\documentclass[amsmath,aps,prd,showpacs,a4paper,10pt,twocolumn]{revtex4}

 \usepackage{epsf}
 \usepackage{graphicx}    

\begin{document}

 \newcommand{\be}[1]{\begin{equation}\label{#1}}
 \newcommand{\ee}{\end{equation}}
 \newcommand{\bea}{\begin{eqnarray}}
 \newcommand{\eea}{\end{eqnarray}}
 \def\disp{\displaystyle}

 \def\gsim{ \lower .75ex \hbox{$\sim$} \llap{\raise .27ex \hbox{$>$}} }
 \def\lsim{ \lower .75ex \hbox{$\sim$} \llap{\raise .27ex \hbox{$<$}} }



 \title{\Large \bf Neutron Star as a Mirror for Gravitational Waves}

 \author{Hao~Wei\,}
 \thanks{Corresponding author;\ email:\ haowei@bit.edu.cn}

 \author{Da-Chun~Qiang\,}
 \author{Zhong-Xi~Yu\,}
 \author{Hua-Kai~Deng\,}

 \affiliation{\vspace{1mm} 
 School of Physics, Beijing Institute of Technology, Beijing 100081, China}

 \begin{abstract} 
 Gravitational wave (GW) has become one of the most active fields in
 physics and astronomy since the first direct detection of GW event
 in 2015. As is well known, multiple images of GW events are possible
 through the gravitational lenses. Here, we propose a novel mirror imaging
 mechanism for GW events different from the gravitational lens. In the
 literature, the superconductor was predicted to be highly reflective mirror
 for GWs. It is well known that neutron stars exhibit superconductivity and
 superfluidity. In this work, we predict that there are two types of GW
 mirror imaging phenomena caused by the neutron star located in Milky Way
 or the same host galaxy of GW source, which might be detected within a
 life period of man (namely the time delay $\Delta t$ can be a few years to
 a few tens of years). It is expected to witness this predicted GW mirror
 imaging phenomenon in the near future. In the long term, the observations
 of this novel GW mirror imaging phenomenon might help us to find numerous
 neutron stars unseen by other means, and learn more about the complicated
 internal structures of neutron stars, as well as their equations of state.
 \end{abstract}

 \pacs{04.30.-w, 04.30.Tv, 97.60.Jd \hfill arXiv:1911.04201}

 \maketitle


 \renewcommand{\baselinestretch}{1.0}



Gravitational wave (GW) is a long-standing prediction of general relativity
 (GR) since 1916 \cite{Einstein:1916cc,Einstein:1918btx,Einstein:1937qu,
 Cervantes-Cota:2016zjc}. The debate about the physical
 reality of GW was mainly settled during the Chapel Hill conference in
 1957~\cite{Saulson:2010zz,Cervantes-Cota:2016zjc}. In 1974, Hulse and
 Taylor discovered the first indirect evidence for the existence of GW
 from the binary pulsar PSR B1913+16~\cite{Taylor:1979zz,Taylor:1982zz,
 Lorimer:2008se}, and hence they earned the 1993 Nobel Prize.
 In 2015, the LIGO/Virgo Collaboration made the first direct observation
 of GW event (GW150914) from a binary black hole
 merger~\cite{Abbott:2016blz,TheLIGOScientific:2016src}, and then earned
 the 2017 Nobel Prize. From the first multi-messenger observations of
 a binary neutron star merger (GW170817) \cite{TheLIGOScientific:2017qsa,
 GBM:2017lvd}, the speed of GW was confirmed to be the speed of light
 $c$~\cite{Monitor:2017mdv}. Nowadays, GW becomes one of the most active
 fields in physics and astronomy.

As is well known, most objects (e.g. earth) are nearly transparent to GWs.
 On the other hand, the gravitational lenses (e.g. massive galaxies) can
 deflect GW as well as light~\cite{Lens1,Lens2}. Therefore, multiple images
 of GW events are possible through the gravitational lenses. Here, we
 propose a novel mirror imaging mechanism for GW events different from the
 gravitational lens.

We note that in \cite{Minter:2009fx,Chiao:2009tn,Chiao:2007pe,Chiao:2017rfe}
 the superconducting film~was predicted to be highly reflective mirror for
 GWs (see also e.g.~\cite{Quach:2015qwa,Inan:2017qdt,Inan:2017ixx}).
 Following~\cite{Minter:2009fx,Chiao:2009tn,Chiao:2007pe,Chiao:2017rfe,
 Quach:2015qwa,Inan:2017qdt,Inan:2017ixx}, here we present a physical
 picture for the possible GW mirror reflection in a superconductor. As
 is well known, the main effect of a GW is making the particles follow
 the distortion in spacetime and then float (freely fall). In a
 superconductor, negatively charged Cooper pairs will be formed according to
 the Bardeen-Cooper-Schrieffer (BCS) theory of superconductivity. The Cooper
 pairs in the ground state are in an exactly zero-momentum eigenstate. So,
 their positions are completely uncertain, i.e. their trajectories are
 completely delocalized, due to the well-known Heisenberg uncertainty
 relation for momentum and position. Note that the quantum delocalization of
 Cooper pairs is protected from the localizing effect of decoherence by the
 BCS energy gap. As a result, Cooper pairs cannot freely fall along with the
 positively charged ions and normal electrons. So, in the presence of GW,
 Cooper pairs undergo non-geodesic motion relative to the geodesic motion of
 its ionic lattice. In other words, Cooper pairs in a superconductor cannot
 respond at all to the passage of GW, in contrast to the positive ions. This
 non-geodesic motion leads to the existence of mass and charge supercurrents
 inside a superconductor. The generation of supercurrents by GW has an
 important consequence, i.e., the electrical polarization of
 the superconductor. The resulting separation of oppositely signed charges
 leads to a huge Coulomb force that strongly opposes the tidal force of the
 incoming GW. So, this GW will be expelled and then reflected. This effect
 in a superconductor is known as the ``\,Heisenberg-Coulomb effect\,''. We
 refer to \cite{Minter:2009fx,Chiao:2009tn,Chiao:2007pe,Chiao:2017rfe,
 Quach:2015qwa,Inan:2017qdt,Inan:2017ixx} for more details.

We stress that the Heisenberg-Coulomb \mbox{effect} {\em cannot occur
 in the normal matters} \cite{Minter:2009fx,Chiao:2009tn,Chiao:2007pe,
 Chiao:2017rfe,Quach:2015qwa,Inan:2017qdt,Inan:2017ixx}, because no Cooper
 pairs can be formed in the normal matters. According to the equivalence
 principle, all particles in the normal matters freely fall along geodesics
 (namely decoherence-induced trajectories). As a result, the normal matters
 cannot energetically interact with GWs, i.e., they are almost transparent
 to GWs. The GW reflection is {\em impossible for the normal matters}.
 However, as mentioned above, the quantum delocalization of Cooper pairs
 (which undergo non-geodesic motion) is the key to make difference in a
 superconductor. Thus, the reflection of GWs becomes possible
 {\em only for the superconductors}.

Let us further discuss some key details following
 \cite{Minter:2009fx,Chiao:2009tn,Chiao:2007pe,Chiao:2017rfe}. As is
 well known, in the limit of weak gravitational field and non-relativistic
 matter, Einstein's field equations can be approximately recast as the
 gravitational Maxwell-like equations~\cite{Braginsky:1976rb,Wald:1984rg,
 Harris:1991,Clark:2000ff}. In such a formalism, one can consider the
 reflectivity of the incident GW, {\em in close analogy with
 the electromagnetic (EM) case}. The characteristic gravitational impedance
 $Z_G=\sqrt{\mu_G/\epsilon_G}=4\pi G/c\sim {\cal O}(10^{-18})$ SI units
 from the gravitational Maxwell-like equations is much less than the EM
 counterpart $Z_0=\sqrt{\mu_0/\epsilon_0}\,$. \mbox{All normal (classical)}
 matters have extremely high levels of \mbox{dissipation} compared to
 $Z_G$, and hence they are \mbox{inevitably} very poor reflectors of GWs. In
 contrast, superconductors are effectively \mbox{dissipationless} at
 temperatures near absolute zero because of their quantum mechanical
 nature. The fact that \mbox{superconductor's} effectively {\em zero}
 impedance can be much less than the very small $Z_G$ allows it to reflect
 an incoming GW, similar to a low-impedance connection at the end of a
 transmission line can reflect an incoming EM wave. One can derive the
 reflectivity ${\cal R}_{\rm G}$ for an incoming GW {\em in close analogy
 with the EM case}. However, the Heisenberg-Coulomb effect makes a crucial
 difference, i.e. {\em an enormously enhanced interaction between the
 incoming GW and superconductor}. The magnitude of the enhancement is
 given by the ratio of the electrical force to the gravitational force
 between two electrons, i.e. $(q^2 Z_0)/(m^2 Z_G)=e^2/(4\pi\epsilon_0 G
 m_e^2)\simeq 4.2\times 10^{42}$. It makes $\Sigma\equiv(\sigma_{1,\,G}/
 \sigma_{2,\,G})^2=(\omega/\omega_c)^2\ll 1$ when the angular frequency
 $\omega\ll\omega_c$\,, where $\omega_c$ is a characteristic angular
 frequency associated with the modified plasma and resonance frequencies of
 a superconducting film. Typically, $\Sigma\sim 10^{-16}$ for $\omega_c\sim
 10^{16}\,{\rm rad/s}$ and $\omega\sim{\cal O}(1)\;{\rm rad/s}$. In this
 case, the GW reflectivity ${\cal R}_{\rm G}=[\,(1+\Sigma)^2+
 \Sigma\,]^{-1}\to 1$. So, the incoming GW at angular frequency
 $\omega\ll\omega_c$ will be almost $100\%$ reflected. We refer
 to \cite{Minter:2009fx,Chiao:2009tn,Chiao:2007pe,Chiao:2017rfe} for the
 detailed derivations.

However, it was found in \cite{Inan:2017qdt,Inan:2017ixx,InanPhD} that
 some technical details of \cite{Minter:2009fx,Chiao:2009tn,Chiao:2007pe,
 Chiao:2017rfe} are faulty (we thank the referee for pointing out this
 issue). Actually, the reflectivity ${\cal R}_{\rm G}$ in
 \cite{Minter:2009fx,Chiao:2009tn,Chiao:2007pe,Chiao:2017rfe} was obtained
 by using vector field equations, vector coupling rule, and
 vector constituent equation, which do not apply to GWs.
 Thus, the gravitational impedance is associated with local oscillating
 gravito-electromagnetic fields rather than GWs \cite{Inan:2017qdt,
 Inan:2017ixx,InanPhD}. Instead, to obtain the impedance associated with
 GWs, one should use tensor field equations, tensor coupling rule, and
 tensor constituent equation (we thank the referee for pointing out this
 issue). They are given by \cite{Inan:2017qdt,Inan:2017ixx,InanPhD}
 \bea
 &\disp -\frac{1}{c^2}\,\partial_t^2 h_{ij}^{TT}+\nabla^2 h_{ij}^{TT}
 = -2\kappa\, T_{ij}^{TT}\,,\label{eq1}\\[0.3mm]
 & \pi^2=p^2+h_{ij}^{TT} p^i p^j\,,\label{eq2}\\[0.9mm]
 & T_{ij}=\mu_G E_{ij}\,,\label{eq3}
 \eea
 respectively, where $h_{ij}^{TT}$ is the transverse-traceless GW field,
 $T_{ij}^{TT}$ is the transverse-traceless stress tensor, $\pi^i$ is the
 kinetic momentum (please do not confuse it with the constant $\pi$),
 $p^i$ is the canonical momentum, $E_{ij}=-\partial_t h_{ij}^{TT}$ is a
 gravito-electric tensor field, $\mu_G$ is the gravitational conductivity,
 and $\kappa=8\pi G/c^4$. In this formalism, one can find that the GW
 impedance is given by \cite{Inan:2017qdt,Inan:2017ixx,InanPhD}
 \be{eq4}
 Z_G^{\,(SC)}=\frac{4\pi G/c}{\sqrt{1-2\,c^2\kappa\,\mu_G/\omega^2\,}}\,.
 \ee
 And then, the GW reflectivity
 reads \cite{Inan:2017qdt,Inan:2017ixx,InanPhD}
 \be{eq5}
 {\cal R}_{\rm G}=\left[1+\left(\frac{c^2}{2}
 \frac{\omega}{\mu_G Z_G^{\,(SC)} d}\right)^2\,\right]^{-1}\,,
 \ee
 where $d$ is the thickness of a thin superconducting film (by ``\,thin\,''
 we mean small relative to the wavelength of GW). We refer to
 \cite{Inan:2017qdt,Inan:2017ixx,InanPhD} for the detailed derivations.

Due to historical reasons, mainly the GWs at microwave frequencies were
 considered in~\cite{Minter:2009fx,Chiao:2009tn,Chiao:2007pe,Chiao:2017rfe,
 Quach:2015qwa,Inan:2017qdt,Inan:2017ixx}, far before the first direct GW
 detection by LIGO. Unfortunately, the experiment for laboratory-scale
 superconducting mirror of GWs is actually difficult to achieve on earth.
 For a usual superconductor, it was found in
 \cite{Inan:2017qdt,Inan:2017ixx} that $\mu_G\sim 10^8\,{\rm J/m^3}$,
 $Z_G^{\,(SC)}\simeq Z_G^{\,(vacuum)}=4\pi G/c$, and then
 ${\cal R}_{\rm G}\sim 10^{-36}$, which implies effectively
 there is no reflection. Obviously, from Eq.~(\ref{eq5}), substantial
 reflection requires
 \be{eq6}
 \frac{c^2}{2}\frac{\omega}{\mu_G Z_G^{\,(SC)} d}\ll 1\,.
 \ee
 Further, one can take the limit when a superconductor behaves
 like classical matter, and then $\mu_G=\rho v^2/2$ which is
 the kinetic energy density of the material. One can even use the upper
 bound $\mu_G\sim \rho c^2$. From Eq.~(\ref{eq6}), the mass density
 required for substantial reflection of GWs at microwave frequency is
 $\rho\sim 10^{29}\,{\rm kg/m^3}$. It is much greater than the mass density
 of neutron stars $\rho\sim 10^{17}\,{\rm kg/m^3}$. So, the reflection
 of GWs at {\em microwave} frequency is completely negligible (we thank
 the referee for pointing out the above issues).

From Eq.~(\ref{eq6}), the way out is to consider a lower frequency $\omega$
 (less than the microwave frequency) and a very large mass density $\rho$ as
 possible, which cannot be found in terrestrial laboratories.

Let us turn eyes to the sky. It is well known that neutron stars exhibit
 superconductivity and superfluidity (see e.g.~\cite{Lattimer:2004pg,
 Baym:1978jf,Weber:2004kj,Page:2010aw,Baldo:2002ju}). Due to the huge
 luminosity distances of GW sources (namely ${\cal O}(10^2)$ Mpc, or,
 ${\cal O}(10^8)$ light years, see e.g.~\cite{Abbott:2016blz,
 TheLIGOScientific:2016src,TheLIGOScientific:2017qsa,GBM:2017lvd,LIGOdect,
 LIGOScientific:2018mvr}) and the small size of neutron stars (namely
 ${\cal O}(10)$ kilometers, see e.g.~\cite{Lattimer:2004pg,Baym:1978jf}),
 it is safe to ignore the spherical shapes of neutron stars and their
 complicated internal structures, and hence one can simply regard them
 as superconducting \mbox{films}. For neutron stars with
 $\rho\sim 10^{17}\,{\rm kg/m^3}$ and $d\sim 10^4\,{\rm m}$, one can
 find from Eq.~(\ref{eq6}) that substantial reflection is only possible
 for GWs at angular frequency $\omega\ll 10^6\,{\rm rad/s}$ (we thank
 the referee for pointing out this issue). Note that GWs could be detected
 by LIGO/Virgo are at frequencies 1\,{\rm Hz} to 100\,{\rm Hz}, and hence
 the angular frequency $\omega\lesssim 700\,{\rm rad/s}\ll
 10^6\,{\rm rad/s}$. In this case, the \mbox{GW} reflectivity
 ${\cal R}_{\rm G}\to 100\%$. Therefore, neutron stars can play the role
 of plane mirrors for GWs at low frequencies.

In principle, the positions of neutron stars will be perturbed
 when GWs go to them. But, such kind of perturbations can also be safely
 ignored with respect to the huge luminosity distances of GW sources.

By estimating the number of stars that have undergone supernova
 explosions, there are thought to be around $10^8$ neutron stars in
 Milky Way \cite{Camenzind:2007book}. But most of them are old, cold and
 hence almost undetectable. It is easy to imagine that numerous neutron
 stars are distributed over the universe. Although most of them cannot
 be seen by other means, neutron stars might manifest themselves in
 the mirror reflection of GWs.


 \begin{center}
 \begin{figure}[tb]
 \centering
 \vspace{-3mm}  
 \includegraphics[width=0.8\columnwidth]{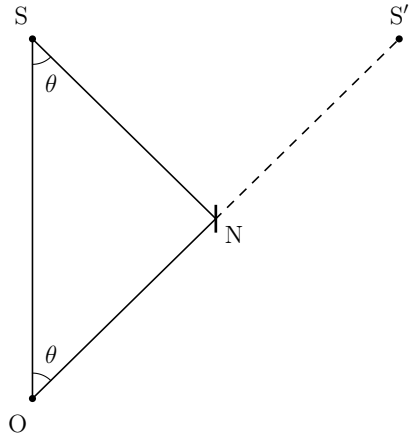}
 \caption{\label{fig1} A simple and naive case of neutron star
 as a mirror for GWs. The plot is not to scale. See the text for details.}
 \end{figure}
 \end{center}



 \begin{center}
 \begin{figure*}[tb]
 \centering
 \vspace{-9mm}  
 \includegraphics[width=0.8\textwidth]{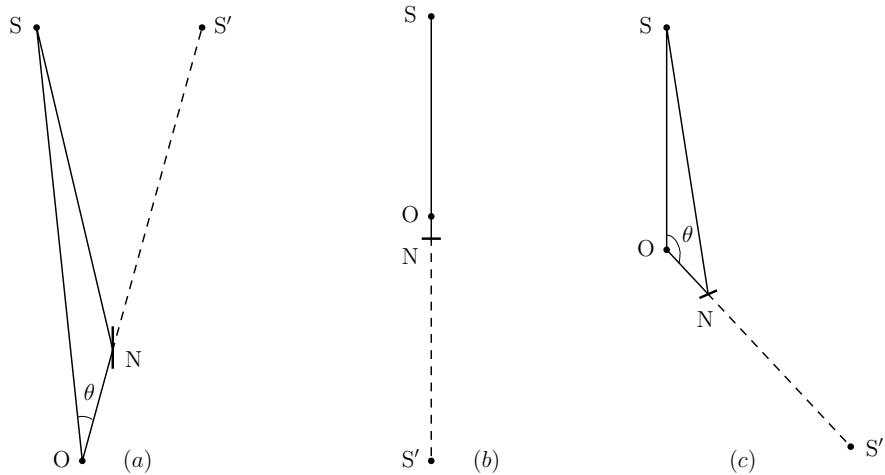}
 \caption{\label{fig2} The cases of neutron star as a mirror for GWs with
 luminosity distances $d_{\rm ON}\ll d_{\rm SN}$, namely the neutron star N
 is much closer to the observer O on earth. The plots are not to scale. See
 the text for details.}
 \end{figure*}
 \end{center}


 \vspace{-18mm}  

At first, let us consider a simple and naive case given in Fig.~\ref{fig1}
 (not to scale). We suppose a transient GW event ``\,S\,'' is directly
 detected by the observer ``\,O\,'' on earth, while GW goes through the
 luminosity distance $d_{\rm OS}$. A neutron star ``\,N\,''~\cite{Nisnot}
 as a plane mirror for GWs located on the vertex of the isosceles triangle
 SNO can reflect the GW from S going to it, and hence the observer O
 will detect a secondary GW signal later. Thus, O sees a mirror image
 ``\,$\rm S^\prime$\,'' of the GW source S. The luminosity distances
 satisfy $d_{\rm OS^\prime}\cos\theta=d_{\rm OS}\,$. The intensity of the
 image $\rm S^\prime$ divided by the intensity of S equals to
 $(d_{\rm OS^\prime}/d_{\rm OS})^{-2}=(\cos\theta)^2$. For example,
 $(\cos\theta)^2=3/4$ and $1/2$ even for $\theta=30^\circ$ and $45^\circ$,
 respectively. Thus, the second GW signal is still detectable even if the
 angle $\theta$ is fairly large. Note that the allowed large angle $\theta$
 between S and $\rm S^\prime$ makes the mirror phenomenon different from
 the gravitational lens whose Einstein angle is typically small. It is
 of interest to calculate the interval between the arrival times of GW
 signals S and ${\rm S}^\prime$, which is given by $\Delta t
 =t_{\rm S^\prime}-t_{\rm S}=(t_{\rm S^\prime}/t_{\rm S}-1)\,t_{\rm S}
 =(d_{{\rm OS^\prime}}/d_{\rm OS}-1)\,t_{\rm S}=
 ((\cos\theta)^{-1}-1)\,t_{\rm S}\,$. Unfortunately, for a not so small
 angle $\theta$, the arrival time interval $\Delta t$ is on the order of
 $t_{\rm S}\sim{\cal O}(10^8)$ years. This is far beyond the patience.
 If we hope to detect both GW signals S and $\rm S^\prime$ within a life
 period of man (namely a few tens of years), the angle $\theta$ must
 be very small. Noting that $\Delta t \simeq \theta^2\,t_{\rm S}/2$ for
 $\theta\to 0$, we find that $\theta\lesssim 10^{-3}$ or $10^{-4}$ is
 required by $\Delta t\lesssim 10^2$ years or $10$ years, respectively.
 This means that the host galaxy of neutron star N lies in the middle of
 the line of sight from O to S. It will play the role of gravitational
 lens. Thus, the mirror imaging phenomenon will be destroyed by the host
 galaxy of neutron star N in this simple and naive case.

Clearly, in the previous case it is too special that S, N, O form an
 isosceles triangle. Noting that any observer O on the extension line
 of $\rm S^\prime N$ can always see the mirror image ${\rm S}^\prime$,
 we can consider the cases with $d_{\rm ON}\ll d_{\rm SN}$, i.e. the
 neutron star N is much closer to O, as shown in Fig.~\ref{fig2} (not to
 scale). In fact, N can be located in the same host galaxy of O (namely
 Milky Way). As is well known, the diameter of Milky Way is
 $1.5\sim 2\times 10^5$ light years~\cite{LopezCorredoira:2018}, and hence
 $d_{\rm ON}\lesssim 2\times 10^5$ light years. However, the neutron star N
 cannot be too close to earth (O), otherwise the humankind and other lives
 on earth will all be killed by it. To our best knowledge, the nearest
 neutron star to earth is claimed to be Calvera (1RXS J141256.0+792204) at
 the distance around $250\sim 1000$ light years~\cite{Rutledge:2007fw}.
 Other two closest known neutron stars are RX J1856.5$-$3754
 and PSR J0108$-$1431, which are both about 400 light years
 from earth~\cite{Posselt:2008ka}. So, we take $d_{\rm ON}\gtrsim 250$
 light years. There are mainly three types of configurations for O and
 N, as shown in Fig.~\ref{fig2}. The case (2b) is fairly special, and it
 becomes possible because earth is nearly transparent to GWs. The mirror
 image $\rm S^\prime$ is in the opposite direction of the GW source S.
 In this case, the interval between the arrival times of GW signals S
 and ${\rm S}^\prime$ is $\Delta t=2d_{\rm ON}/c\gtrsim 500$ years. We turn
 to the case (2c) with the angle $\theta\geq 90^\circ$. In this case,
 $\Delta t=(d_{\rm SN}+d_{\rm ON}-d_{\rm OS})/c > d_{\rm ON}/c\gtrsim 250$
 years. Thus, both the arrival time intervals $\Delta t$ of cases (2b) and
 (2c) are far beyond a life period of man (namely a few tens of years). The
 last hope lies in the case (2a) with the angle $\theta< 90^\circ$.
 Obviously, $\Delta t=(d_{\rm SN}+d_{\rm ON}-d_{\rm OS})/c \to 0$ if
 $\theta\to 0$. However, we should avoid $\theta\to 0$, otherwise the GW
 from S might be reflected before it could reach the observer O. In fact,
 a large $\theta$ is possible in the case (2a). Using the law of cosines
 to the triangle SON, \mbox{we have} $\cos\theta=(d_{\rm OS}^2
 +d_{\rm ON}^2-d_{\rm SN}^2)/(2d_{\rm OS}d_{\rm ON})=\\[0.7mm]
 (d_{\rm OS}^2+d_{\rm ON}^2-
 (d_{\rm OS}-d_{\rm ON}+c\Delta t)^2)/(2d_{\rm OS}d_{\rm ON})\simeq\\[0.6mm]
 1-c\Delta t/d_{\rm ON}$ for $d_{\rm ON}\ll d_{\rm OS}$ and
 $c\Delta t\ll d_{\rm OS}\,$. So, we obtain $\Delta t\simeq\left(1-
 \cos\theta\right)d_{\rm ON}/c\,$. If $d_{\rm ON}\simeq 250$ light years,
 $\Delta t\simeq 0.04$, $0.95$, $3.8$, $8.5$, $15.1$, $33.5$, $47.7$ and
 $73.2$ years for $\theta=1^\circ$, $5^\circ$, $10^\circ$, $15^\circ$,
 $20^\circ$, $30^\circ$, $36^\circ$ and $45^\circ$, respectively. If
 $d_{\rm ON}\simeq 1000$ light years, $\Delta t\simeq 0.15$, $3.8$, $15.2$,
 $34.1$ and $60.3$ years for $\theta=1^\circ$, $5^\circ$, $10^\circ$,
 $15^\circ$ and $20^\circ$, respectively. Thus, we can see the mirror
 image $\rm S^\prime$ within a life period of man (namely the time delay
 $\Delta t$ can be a few years to a few tens of years) while the angle
 $\theta$ can still be fairly large. On the other hand, the intensity of
 the image ${\rm S}^\prime$ divided by the intensity of S equals to
 $(d_{\rm OS^\prime}/d_{\rm OS})^{-2}=(1+c\Delta t/d_{\rm OS})^{-2}\simeq 1$
 for $c\Delta t\ll d_{\rm OS}\,$. So, we will see two GW signals S and
 $\rm S^\prime$ with almost same intensities and wave forms.


 \begin{center}
 \begin{figure*}[tb]
 \centering
 \vspace{-8mm}  
 \includegraphics[width=0.8\textwidth]{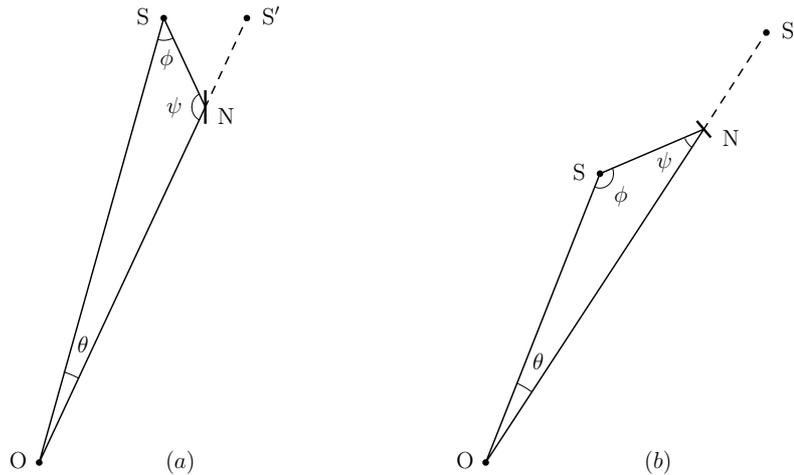}
 \caption{\label{fig3} The cases of neutron star as a mirror for GWs with
 luminosity distances $d_{\rm SN}\ll d_{\rm ON}$, namely the neutron star N
 is much closer to the GW source S. The plots are not to scale. See the
 text for details.}
 \end{figure*}
 \end{center}


 \vspace{-10mm}  

Similarly, we can also consider the cases with $d_{\rm SN}\ll d_{\rm ON}$,
 namely the neutron star N is much closer to the GW source S, as shown in
 Fig.~\ref{fig3} (not to scale). In fact, N can be located in the same host
 galaxy of S. Noting that the remnants of most GW events are black holes
 which can also swallow GWs, in Fig.~\ref{fig3} there is no counterpart to
 the case (2b). Since we need not to care the alien lives, the neutron star
 N can be very close to S, unlike the cases in Fig.~\ref{fig2}. In fact,
 $d_{\rm SN}$ can be a few light years, or a few tens of light years.
 Because $d_{\rm SN}\ll d_{\rm ON}\sim d_{\rm OS}$, it is easy to see that
 the angle $\theta$ must be very close to 0 (on the order of
 $d_{\rm SN}/d_{\rm OS}\sim 10^{-8}$ to $10^{-7}$). Thus, from the viewpoint
 of the observer O on earth, the mirror image $\rm S^\prime$ is almost in
 the same direction of the GW source S. But this does not mean that the
 neutron star N must lie in the line OS or its extension line, from the
 local viewpoints of S and N, as will be shown below. In the case (3a) with
 the angle $\phi<90^\circ$, following the similar derivations in the
 previous case (2a), we have $\cos\phi\simeq 1-c\Delta t/d_{\rm SN}$.
 Clearly, the angle $\phi$ can be large if $c\Delta t$ is comparable to
 $d_{\rm SN}$. So, the neutron star N does not lie in the line OS actually
 (although $\theta\to 0$ due to the huge $d_{\rm OS}$). Because
 $c\Delta t\simeq\left(1-\cos\phi\right)d_{\rm SN}$, we can see the mirror
 image $\rm S^\prime$ within a life period of man (namely $\Delta t$ can be
 a few years to a few tens of years) if the neutron star N is close enough
 to the GW source S while the angle $\phi$ can still be fairly large. Note
 that even if N is not so close to S, $\Delta t$ can still be a few years to
 a few tens of years for a suitable and large $\theta$ (see the case (2a)).
 Similarly, in the case (3b) with the angle $\phi\geq 90^\circ$ (and hence
 the angle $\psi<90^\circ$), we have $c\Delta t\simeq\left(1+\cos\psi\right)
 d_{\rm SN}$. The angle $\psi$ can be large, so that the neutron star N does
 not lie in the extension line of OS actually (although $\theta\to 0$ due to
 the huge $d_{\rm OS}$). Again, we can see the mirror image $\rm S^\prime$
 within a life period of man (namely $\Delta t$ can be a few years to a few
 tens of years) if the neutron star N is close enough to the GW source S
 while the angle $\psi$ can still be fairly large. On the other hand,
 in both cases (3a) and (3b), the intensity of the image
 ${\rm S}^\prime$ divided by the intensity of S equals to
 $(d_{\rm OS^\prime}/d_{\rm OS})^{-2}=(1+c\Delta t/d_{\rm OS})^{-2}\simeq 1$
 for $c\Delta t\ll d_{\rm OS}$. So, we will see two GW signals S and
 $\rm S^\prime$ with almost same intensities and wave forms.

Besides the three cases shown in Figs.~\ref{fig1}--\ref{fig3} (not to
 scale), the rest is the case with $d_{\rm SN}$ comparable to $d_{\rm ON}$.
 It is easy to see that this case is quite similar to the simple and
 naive case in Fig.~\ref{fig1}, and hence is also inviable. It is worth
 noting that the main reason for the failure of the cases with $d_{\rm SN}$
 comparable to $d_{\rm ON}$ (including the simple and naive case in
 Fig.~\ref{fig1}) is either $\Delta t\sim t_{\rm S}\sim{\cal O}(10^8)$ years
 if the angle $\theta$ is not so small, or the host galaxy of neutron star
 N must play the role of gravitational lens to destroy the mirror imaging
 phenomenon if the angle $\theta$ is very close to 0. In fact, the latter
 might be evaded for an isolated neutron star N without a host galaxy,
 which might be a wandering star in the intergalactic space been kicked
 out of its original position due to some unknown reasons. However, this
 must be extremely rare, and hence we do not consider this possibility here.

In summary, we predict that there are two types of GW mirror imaging
 phenomena caused by the neutron star located in Milky Way (case (2a))
 or the same host galaxy of GW source (cases (3a) and (3b)), which might
 be detected within a life period of man (namely the time delay $\Delta t$
 can be a few years to a few tens of years). In both types of GW mirror
 imaging phenomena, we will see two GW signals S and $\rm S^\prime$ with
 almost same intensities and wave forms. The separate angle $\theta$
 between the GW source S and its mirror image ${\rm S}^\prime$ can be
 fairly large in case (2a), while $\theta$ should be fairly close to 0
 in cases (3a) and (3b). Noting that GWs became directly detectable since
 2015, we hope to witness this predicted GW mirror imaging phenomenon in
 the near future.

In fact, the GW mirror imaging phenomenon predicted here is different
 from the well-known gravitational lenses~\cite{Lens1,Lens2} which can
 also image GWs. In the mirror imaging phenomenon, we will see the signal
 from the GW source S directly, and after a time delay $\Delta t$ we will
 see a second GW signal from the unique mirror image ${\rm S}^\prime$.
 On the contrary, in the case of gravitational lens, one will see more
 than (or equal to) two images, or even rings and arcs. In the mirror
 imaging phenomenon might be detected within a life period of man, we
 will see two GW signals S and $\rm S^\prime$ with almost same intensities
 and wave forms. On the contrary, in the case of gravitational lens, the
 images should be magnified (and/or distorted). Also, in the case of
 gravitational lens, the separate angle $\theta$ (Einstein angle) between
 images is very small (usually a few arcseconds, namely
 $\theta\sim 10^{-6}$). But in the mirror imaging phenomenon, the separate
 angle $\theta$ between the GW source S and its mirror image
 ${\rm S}^\prime$ can be fairly large in case (2a). Finally, the time
 delay $\Delta t$ is usually short (a few days to a few years) in the
 case of gravitational lens, while the time delay $\Delta t$ could be
 relatively long (a few years to ${\cal O}(10^8)$ years) in the mirror
 imaging phenomenon.

Note that the GW mirror imaging phenomenon predicted here is
 also different from other exotic phenomena like Poisson-Arago
 spot~\cite{Hongsheng:2018ibg} or intensification~\cite{Halder:2019cmp}
 for GWs.

Clearly, the discussions in the present work are very preliminary. Many
 simplifications were made. For example, we ignored the spherical shape of
 neutron star and its complicated internal structures. Also, we have not
 calculated the realistic GW reflectivity ${\cal R}_{\rm G}$ for an impure
 superconducting solid sphere. On the other hand, the event rate of this
 GW mirror imaging phenomenon was not estimated. However, the present
 preliminary work has clearly shown the key idea of this GW mirror imaging
 phenomenon, and made clear predictions which might be detected in the near
 future. This is very important.

Actually, this GW mirror imaging phenomenon might not only be detected
 in the future, but also be used to understand the past. For example, it
 has been employed in \cite{Wei:2019wxd} to explain the null result of
 searching the electromagnetic counterparts for the first high-probability
 neutron star -- black hole (NSBH) merger LIGO/Virgo GW190814
 \cite{S190814bv}. It is nothing but a GW mirror image of the real NSBH
 merger before 14 September 2015. This could be regarded as a useful
 support to the GW mirror imaging mechanism.

We hope to construct some full and realistic models for this predicted
 GW mirror imaging phenomenon caused by neutron stars in the future
 works. In the long term, the observations of this novel GW mirror imaging
 phenomenon might help us to find numerous neutron stars unseen by other
 means, and learn more about the complicated internal structures of neutron
 stars, as well as their equations of state.

 \vspace{3mm}  



We are grateful to the anonymous referee for the expert and useful comments
 and \mbox{suggestions}, which have significantly helped us to improve this
 work. We also thank Rong-Gen~Cai, Hongwei~Yu, Zong-Kuan~Guo, Puxun~Wu,
 Zhoujian~Cao, Wen Zhao, Hongsheng~Zhang, Bin~Hu, Yi~Zhang, Hai-Nan~Lin,
 Shou-Long~Li, and Shu-Ling~Li for very useful discussions. H.W. thank
 University of Jinan for the kind hospitality. This work was supported
 in part by NSFC under Grants No.~11975046 and No.~11575022.

\renewcommand{\baselinestretch}{1.12}



\begin{thebibliography}{99}
\bibitem{Nisnot}
  Note that the neutron star ``\,N\,'' as a mirror for \mbox{GWs} is
  clearly not the one in the GW source ``\,S\,'' (which might be binary
  neutron star (BNS) or neutron star -- black hole (NSBH)).
\end{thebibliography}
\end{document}